\documentclass[a4paper]{jpconf}
\usepackage{graphicx}
\usepackage{amsmath}
\begin{document}
\title{Exploiting diurnal temperature variations to monitor the growth of tubers}

\author{Marissa Bezemer and Neil Budko}

\address{Numerical Analysis, DIAM, Delft University of Technology, 2628 XE, Delft, The Netherlands}

\ead{n.v.budko@tudelft.nl}

\begin{abstract}
The possibility to use diurnal temperature variations for nondestructive monitoring of growing tubers is investigated by numerically simulating the data collected with a grid of passive thermal sensors placed in the ground and sampled at regular intervals. A qualitative linear imaging algorithm that produces an approximate projected view of the tubers is proposed and an effective inversion method is applied to recover the volume fraction of tubers. In particular, it is shown that a correlation-based cost functional outperforms the usual least-squares metric, although, requiring additional steps to deal with the non-uniqueness of the solution. 
\end{abstract}

\section{Introduction}
Phenotyping and yield prediction of tuberous and root crops such as potatoes, cassava and yam require digging up, analyzing and discarding numerous plants at various times during the season, which is time-consuming, laborious, destructive and wasteful. Attempts to adapt existing non-destructive subsurface imaging techniques, such as ground penetrating radar \cite{konstantinovic_detection_2007}, face obvious difficulties due to the above-ground leaves and branches that deny direct access to the soil surface. If antennas are to be placed above the plants, then one can expect significant reflection and distortion of probing electromagnetic and acoustic signals, which can not be removed by simple subtraction as canopies grow and change with time.

Natural diurnal air temperature variations create temperature ``waves'' that slowly propagate into the subsurface and rapidly decay with depth \cite{lei_improved_2011}. Similar to the electromagnetic waves of the ground-penetrating radar, these temperature waves are distorted by the tubers, e.g., potatoes. The thermal contrast of tubers with respect to a typical soil is high enough even in wet circumstances due to the difference in heat capacity. Therefore a question emerges whether these natural temperature waves can be exploited to visualize and measure the state of tubers at any given time, in particular, their number, size, shape and location. 
Since temperature waves decay very rapidly, it is clear that sensors must be placed sufficiently close to both the ground surface and the tubers. Where exactly should these sensors be placed, what should be their sensitivity and sampling rate are some of the open questions addressed in this paper. Previously, artificially excited thermal waves have successfully been applied for imaging of inhomogeneous solids \cite{tabatabaei_thermal-wave_2009, ghali_comparative_2011}. Natural diurnal temperature waves, however, do not have a broad spectrum and may require a different approach.

In the following sections realistic three-dimensional heat transfer simulations are employed to determine the characteristic temperature ranges and estimate the required spatial/temporal resolution and sensitivity of temperature sensors. A simple imaging algorithm is proposed to visualize the lateral distribution of tubers and an effective inversion method is applied to nondestructively determine the tubers state over the season.

\section{Modeling}
\begin{figure}
\begin{center}
\includegraphics[width=16cm]{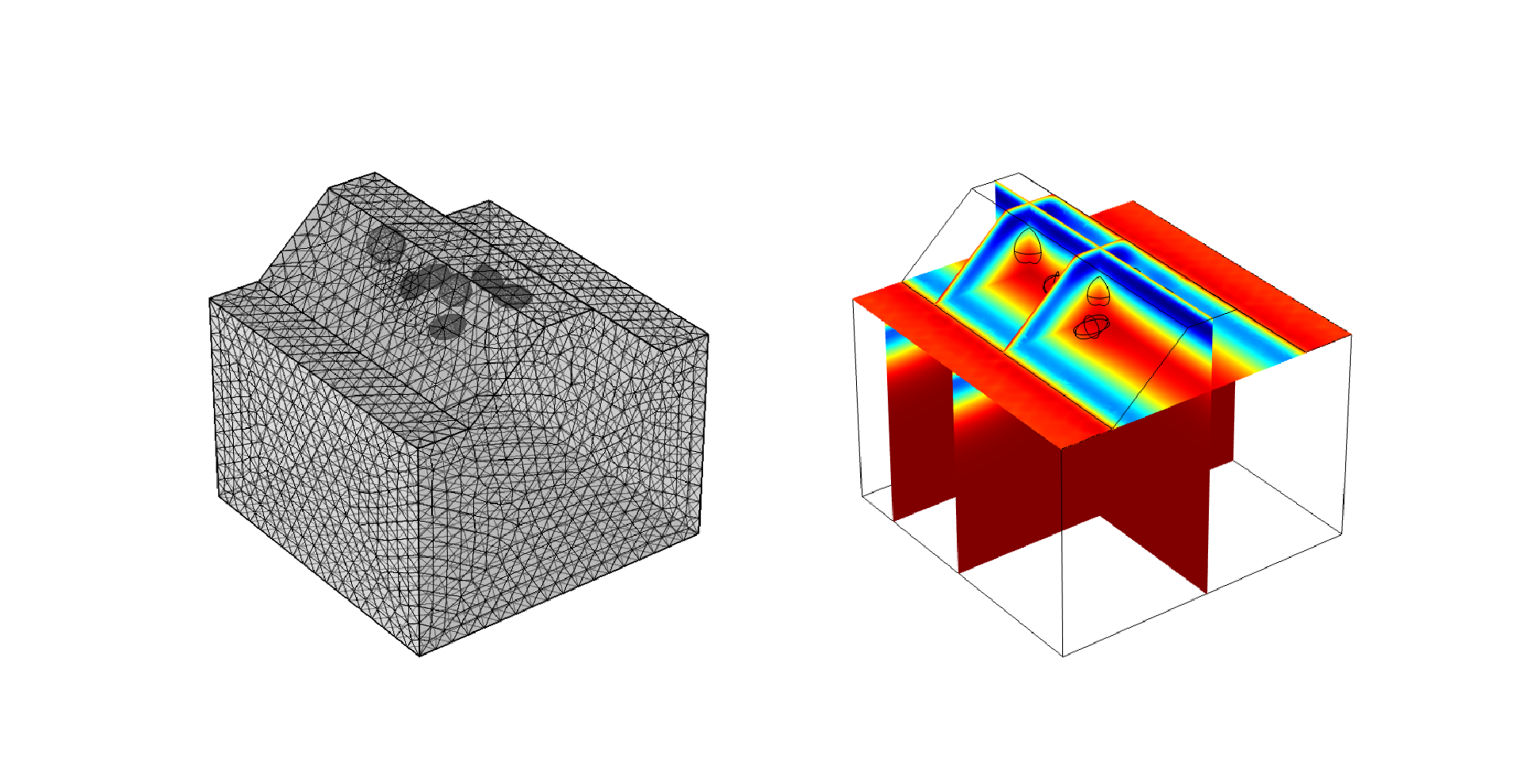}
\end{center}
\caption{\label{Fig1} Three-dimensional configuration with the generated mesh (left) and the computed temperature distribution snapshot showing the subsurface temperature wave (right).}
\end{figure}

The temperature distribution $T({\mathbf x},t)$ in the configuration depicted in Fig.~\ref{Fig1}~(left) is assumed to satisfy the following heat transfer problem:
\begin{align}
\begin{split}
\label{eq:HT}
\rho C_{p}\frac{\partial T}{\partial t} -\nabla\cdot(k\nabla T) = 0,\;\;\;&{\mathbf x}\in \Omega,\;\;\;t\in(0,t_{\rm end}];
\\
T({\mathbf x},0) = T_{\rm amb}(0),\;\;\;&{\mathbf x}\in \Omega;
\\
{\mathbf n}\cdot(k\nabla T) = \epsilon\sigma(T_{\rm amb}^{4}-T^{4}),\;\;\;&{\mathbf x}\in \partial\Omega_{\rm top};
\\
{\mathbf n}\cdot(k\nabla T) = 0,\;\;\;&{\mathbf x}\in \partial\Omega_{\rm bottom};
\\
{\mathbf n}_{1}\cdot(k_{1}\nabla T_{1}) = -{\mathbf n}_{2}\cdot(k_{2}\nabla T_{2});\;\;\;
T_{1}=T_{2},\;\;\;&{\mathbf x}\in \partial\Omega_{\rm sides},
\end{split}
\end{align}
where the last boundary condition is periodic.
The values of the material parameters for soil and potatoes \cite{jong_van_lier_soil_2013, mimouni_estimating_2015, magee_measurement_1995} have been set as in Table~\ref{Table1} and the surface emmissivity was set as $\epsilon = 0.92$ and the simulation time as $t_{\rm end}=192$~h. The ambient air temperature $T_{\rm amb}(t)$ was extracted from historical meteorological data for the period starting on June 1, 2016, in Leeuwarden (Netherlands).

\begin{table}[h]
\caption{\label{Table1}Material parameters.}
\begin{center}
\begin{tabular}{l|l|l|l}
\br
 & $k$, W/(m$\,$K) & $\rho$, kg/$\text{m}^{3}$ & $C_{p}$, J/(kg$\,$K)\\
\mr
Soil & 0.3 & 1300 & 800\\
Potato & 0.56 & 1079 & 4036\\
\br
\end{tabular}
\end{center}
\end{table}

Geometry of the problem mimics a typical ``ridge'' structure where the seed potato is planted in the center of an elevated ridge at the depth of $12.5$~cm from the surface and the new potatoes of different sizes and shapes emerge later at approximately the same average depth and tend to remain within the ridge. Three stages in the growth of new potatoes depicted as horizontal projections of Fig.~\ref{Fig4}~(top) have been modelled. Numerical solution of the heat transfer problem~(\ref{eq:HT}) was obtained with the Finite Element Method (FEM) in the Comsol~Multiphysics~5.4 software package.

A snapshot of the resulting temperature wave can be seen in Fig.~\ref{Fig1}~(right), where on the soil surface the temperature has already reached its daily maximum (red color), whereas the minimum of the preceding night (blue color) is still propagating downwards into the subsurface. It is also clear that the wave diminishes very rapidly with depth as no discernible temperature variations can be seen below the first wave cycle. Nonetheless, it is also clear that the first wave cycle covers the typical potato depths (down to 40~cm).

\begin{figure}
\begin{center}
\includegraphics[width=14cm]{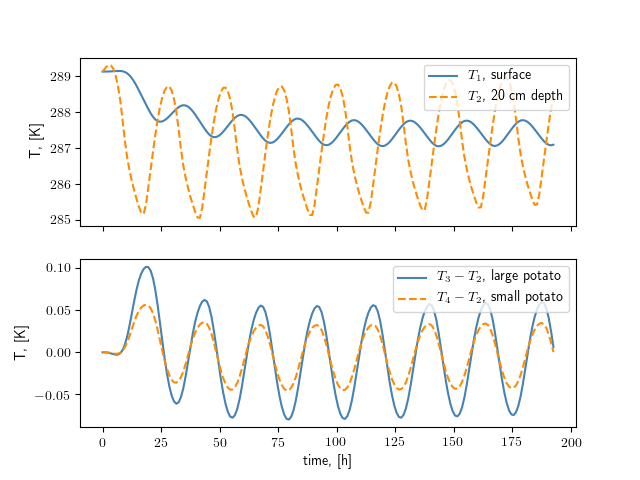}
\end{center}
\caption{\label{Fig2} Top: temperature without potatoes as a function of time; $T_{1}$ -- at the surface, $T_{2}$ -- at $20$~cm depth. Bottom: temperature difference at $20$~cm due to the presence of large and small potatoes (spheres of $10$~cm and $8$~cm diameters with centers at $12.5$~cm depth). Initial parts of the plots (up to $75$~hours) correspond to the transient period in simulations.}
\end{figure}

Figure~\ref{Fig2} shows temperature variations at several points in the configuration. Namely, $T_{1}$ is at the surface above an empty (no potatoes) spot, $T_{2}$ is exactly below the location of $T_{1}$, but at $20$~cm depth, $T_{3}$ and $T_{4}$ are the temperatures measured at $20$~cm depths below a large ($10$~cm diameter) and small ($8$~cm diameter) potatoes, respectively. Note that the bottom plot of Fig.~\ref{Fig2} shows the differences $T_{3}-T_{2}$ and $T_{4}-T_{2}$. The transient period up to $75$~h is an artefact of the wrong initial condition $T({\mathbf x},0)=T_{\rm amb}(0)$ and should be disregarded. The top plot of Fig.~\ref{Fig2} shows an oscillating signal in time, typical for a quasi-harmonic wave, that diminishes in amplitude with depth and undergoes a phase shift. The amplitude of variations reduces from approximately $4$ degrees at the surface to just one degree at $20$~cm depth. As can be seen in the bottom plot of Fig.~\ref{Fig2}, the influence of potatoes on the signal at $20$~cm are in the order of $0.1$ degrees. This means that, if one would like to detect potatoes, then the amplitude resolution should be at least $0.01$~degrees. It is well-known that achieving $0.01$~degree absolute accuracy in temperature measurements is a very difficult task. Therefore, any viable imaging or inversion algorithm should rely on the relative stability and high (amplitude) resolution of the measurements rather than absolute accuracy.

In particular, a relative calibration procedure of temperature sensors should be implemented. For example, if the data are to be collected by a horizontal wire-mesh of thermistors, then all thermistors should be (numerically) calibrated to give the same readings and slopes in an environment with controlled spatially uniform temperature. The high-frequency electronic noise should not be a problem as the data may be collected at hourly intervals, leaving enough time for signal averaging.

\section{Imaging}
\begin{figure}
\begin{center}
\includegraphics[width=14cm]{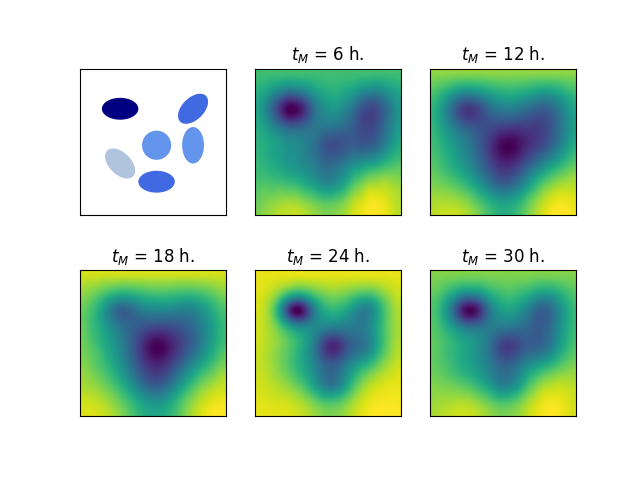}
\end{center}
\caption{\label{Fig3} Top-left: horizontal two-dimensional projection (shadows) of 6 potatoes (see Fig.~\ref{Fig1}) with color coding indicating the distance from the measurement plane (dark means close). Rest: Images $I(x,y,t_{M})$ obtained with eq.~(\ref{eq:Image}) for different $t_{M}$.}
\end{figure}

Comparing the top plot (blue, solid curve) with the curves in the bottom plot of Fig.~\ref{Fig3} one can notice the expected phase shift between the background temperature and the temperature-difference signals. A very basic imaging algorithm can be applied to convert this local phase shift into a two-dimensional image of potatoes. 

Let the input data be collected over a doubly uniform grid $\{(x_{i},y_{j})\,\vert\,i=1,\dots,20;\,j=1,\dots,20\}$
with $2$~cm spatial step located at $20$~cm depth, approximately covering the base of the ridge. Specifically, assuming discrete data measured at $1$-hour intervals $T(x_{i},y_{j},t_{m})$, $m=1,\dots,M$; and the corresponding background temperature $T_{\rm b}(x_{i},y_{j},t_{m})$, $m=1,\dots,M$; measured at the same depth in an area free from potatoes, the image $I(x_{i},y_{j},t_{M})$, can be computed as follows:
\begin{align}
\label{eq:Image}
I(x,y,t_{M}) = L(x,y)\left[ \sum_{m=1}^{M}\left[T(x_{i},y_{j},t_{m})-T_{\rm b}(x_{i},y_{j},t_{m})\right]T_{\rm b}(x_{i},y_{j},t_{m})\right],
\end{align}
where $L(x,y)$ denotes the interpolation operator for locations $(x,y)$ that fall between the grid points $(x_{i},y_{j})$.
Images obtained with different choices of $t_{M}$ are shown in Fig.~\ref{Fig3}, and the $24$-hour interval consisting of $M=25$ time samples per sensor and containing the full temperature-wave cycle appears to provide the best contrast.

\begin{figure}
\begin{center}
\includegraphics[width=14cm]{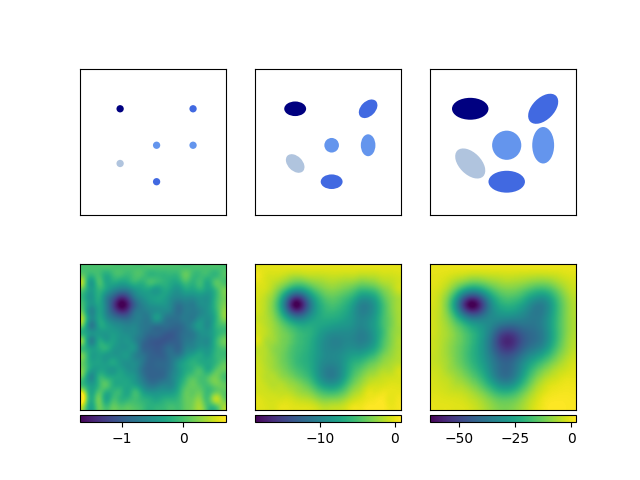}
\end{center}
\caption{\label{Fig4} Images of three growth stages, each based on $24$~h data.}
\end{figure}

The phase shift of the temperature-difference signals with respect to the background signal results in locally negative values of the function $I(x,y,t_{M})$. This negative contrast is, however, not uniquely determined by the size of tubers alone and is also influenced by potato's proximity to the measurement plane. The relatively good lateral resolution of the images stems from the exponential decay of temperature perturbations away from their origin, i.e., tubers, in the present case. At the same time this exponential decay leads to the ``disappearance'' of potatoes situated too far away from the measurement plane. This is a well-known drawback of all imaging algorithms based on diffusive fields \cite{tabatabaei_thermal-wave_2009, ghali_comparative_2011}. Nevertheless, one can easily detect the presence of at least five (out of six) tubers in the images of Fig's~\ref{Fig3}--\ref{Fig4}. 

Despite the difficulty in interpreting the images the algorithm is sensitive enough to capture the emergence and growth of new potatoes as illustrated in Fig.~\ref{Fig4}. The noise in the first image is of numerical origin due to a relatively rough FEM mesh and very weak temperature-difference signals produced by the $2$~cm potato tubers. The colorbars show that the negative contrast increases with the size of tubers. Note that the surfaces of growing tubers move closer to the detector plane as well.

\section{Effective inversion}
Another approach that works with limited and noisy data is called {\it effective inversion} \cite{budko_characterization_1999, budko_electromagnetic_2004, raghunathan_effective_2010}. The basic idea is to fit the data with an easy to interpret low-dimensional model that captures the essence of the object-data relation. When dealing with inhomogeneous targets, such as a group of tubers, the effective model is typically homogeneous and covers the domain where one expects the tubers to be present. The material parameter of this homogeneous model, thermal diffusivity $D=k/(\rho C_{p})$, that produces the best fit to the measured data is denoted as $D_{\rm eff}$ and is called the {\it effective} diffusivity. 

The actual thermal diffusivities of soil and potatoes assumed here, see Table~\ref{Table1}, are $D_{\rm s}=2.88\times 10^{-7}$~$\text{m}^{2}/\text{s}$ and $D_{\rm p}=1.29\times 10^{-7}$~$\text{m}^{2}/\text{s}$, respectively. Hence, the effective diffusivity of the homogeneous model is expected to be somewhere between these values, depending on the amount and sizes of potatoes within the effective domain. During the season, as potatoes emerge and grow, the effective diffusivity $D_{\rm eff}$ will start off as $D_{\rm s}$ and then gradually approach a smaller value closer to $D_{\rm p}$, thus indicating the amount and the rate of growth of potatoes.

\begin{figure}
\begin{center}
\includegraphics[width=16cm]{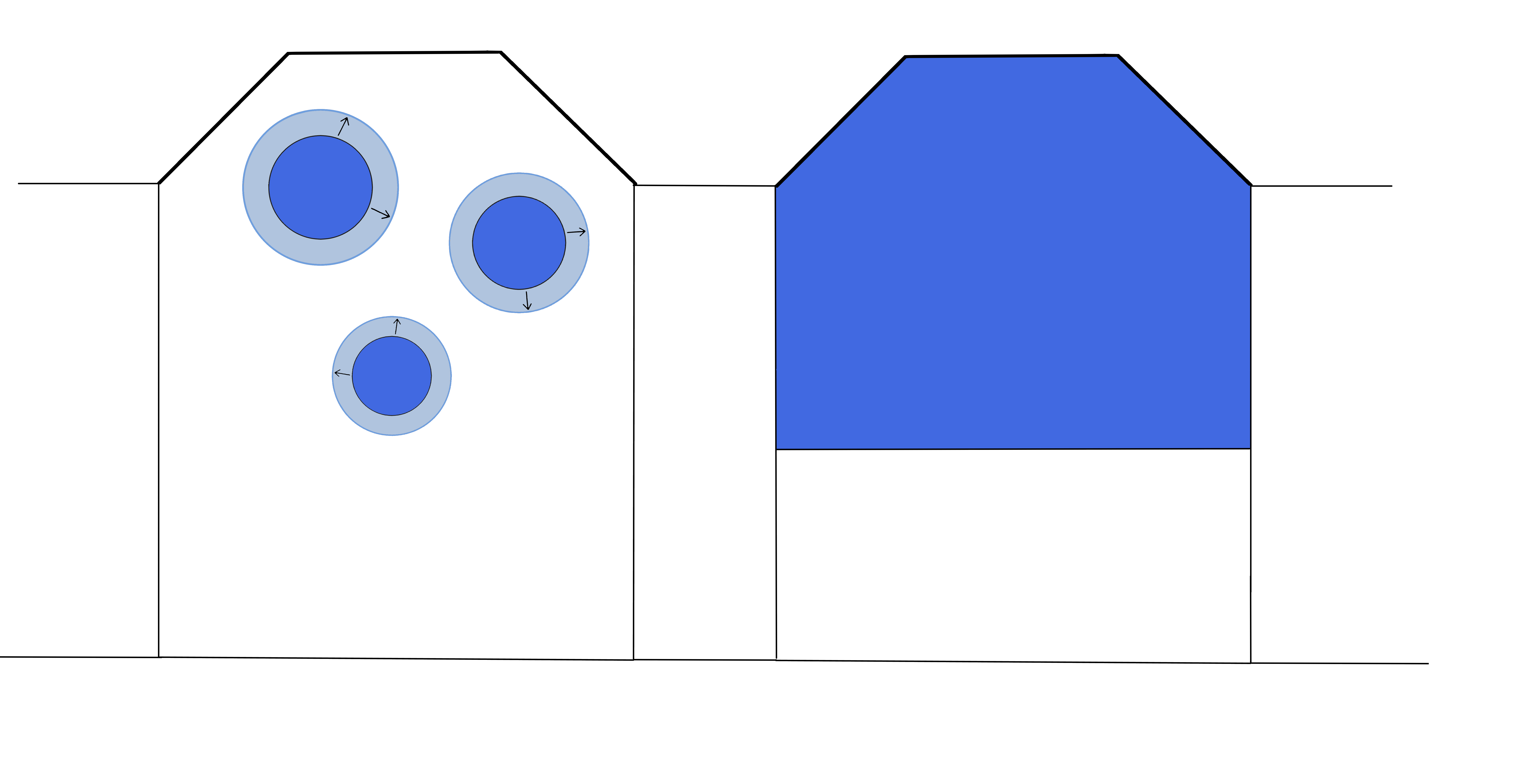}
\end{center}
\caption{\label{Fig5} Schematics of the original (left) and effective (right) two-dimensional models. The growing potatoes are modelled as the time varying diffusivity of a homogeneous effective domain.}
\end{figure}

Numerical experiments in this Section have been performed on the two-dimensional configuration depicted in Fig.~\ref{Fig5}. The growth of potatoes is modelled on a fine temporal scale with the radius of each tuber increasing at the rate of $0.86$~mm/day, reaching the final radius of $6$~cm at the end of the $70$-day season. The boundary conditions and material parameters are the same as in eq.~(\ref{eq:HT}) and Table~\ref{Table1} with the exception of emissivity, which was set to $\epsilon=0.8$. The corresponding heat transfer problem is solved numerically with the Finite-Element Method using the Fenics (dolfin) module in Python. There are $12$ randomly placed potatoes in the particular numerical experiments presented below.

Let $T(x_{i},t_{m})$ be the simulated data generated by the forward heat transfer model of growing potatoes with diffusivity $D_{\rm p}$, where $x_{i}$, $i=1,\dots,N$, are the locations of thermal detectors uniformly placed at $1$~cm intervals along a horizontal $60$~cm line segment situated $45$~cm below the top of the ridge. We denote as $T_{\rm eff}(x_{i},t_{m},D)$ the temperature that would have been measured at the same locations and at the same times if instead of potatoes there would be a homogeneous block with diffusivity $D$ as shown in Fig.~\ref{Fig5}~(right). The diffusivity of the soil $D_{\rm s}$, the ambient temperature $T_{\rm a}(t)$, and the background (no potatoes) temperature $T_{\rm b}(x_{i},t_{m})$ at the detector line are considered to be known and the same in both models. The goal is to match the data $T(x_{i},t_{m})$ and the temperature $T_{\rm eff}(x_{i},t_{m},D)$ by tuning the diffusivity $D$ of the effective model. The value of $D$ that provides the best match is the effective diffusivity $D_{\rm eff}$. Since potatoes grow, the effective diffusivity will depend on time.

\begin{figure}
\begin{center}
\includegraphics[width=13cm]{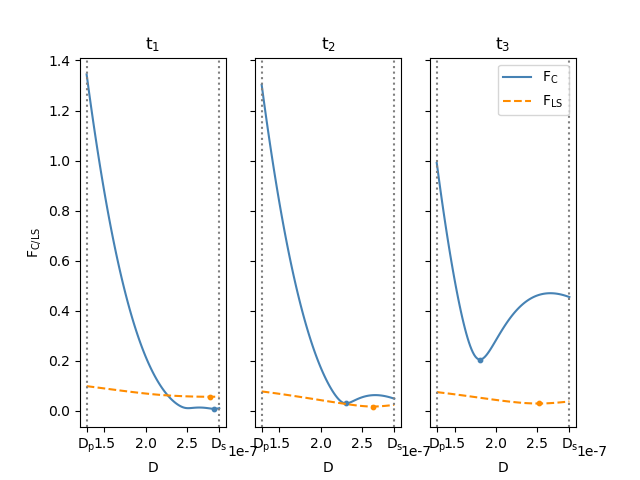}
\end{center}
\caption{\label{Fig6} Shapes of least-squares and correlation-based cost functionals based on $24$-hour data sampled at hourly intervals at three days during the growth season ($t_{1}=3$, $t_{2}=35$, and $t_{3}=70$~days). Actual configuration consists of $12$ circular growing ``potatoes'' randomly placed within the boundaries of the effective domain.}
\end{figure}

Matching data to the model output requires defining a cost functional, which is not trivial in the case where data depend on both space and time. The most natural choice would be the following least-squares functional:
\begin{align}
\label{eq:FuncLS}
F_{\rm LS}\left[D\right] = \frac{\sum_{i=1}^{N}\sum_{m=1}^{M}\left[T(x_{i},t_{m})-T_{\rm eff}(x_{i},t_{m},D)\right]^{2}}{\sum_{i=1}^{N}\sum_{m=1}^{M}\left[T(x_{i},t_{m})-\overline{T}(x_{i})\right]^{2}},
\end{align}
where $\overline{T}(x_{i})=(1/M)\sum_{m=1}^{M}T(x_{i},t_{m})$, which provides for a better normalization. Numerical experiments reveal, however, that the effective diffusivity reconstructed with this functional fails to capture the growth of potatoes. Namely, $D_{\rm eff}$ remains in the neighborhood of the soil diffusivity $D_{\rm s}$ during most of the growing season, see Fig.~\ref{Fig6}. 

As was discussed in the previous Section, phase shift of the temperature signal due to potatoes appears to be rather significant. The following correlation-based functional is aimed at equalizing phase shifts:
\begin{align}
\label{eq:FuncC}
F_{\rm C}\left[D\right] = \frac{\sum_{i=1}^{N}\left[C_{i}(T)-C_{i}(T_{\rm eff})\right]^{2}}{\sum_{i=1}^{N}\left[C_{i}(T)\right]^{2}},
\end{align}
where $C_{i}(T)$ measures by how much the data or the effective model are shifted with respect to the background temperature signal, i.e.,
\begin{align}
\begin{split}
\label{eq:Cdef}
C_{i}(T) &= \frac{\sum_{m=1}^{M}\left[T(x_{i},t_{m})-\overline{T}(x_{i})\right]\left[T_{\rm b}(x_{i},t_{m})-\overline{T}_{\rm b}(x_{i})\right]}{\sigma_{i}(T)\sigma_{i}(T_{\rm b})},
\\
\sigma_{i}(T) &= \left[\sum_{m=1}^{M}\left\vert T(x_{i},t_{m})-\overline{T}(x_{i})\right\vert^{2}\right]^{1/2}.
\end{split}
\end{align}
Figure~\ref{Fig6} demonstrates the shapes of $F_{\rm LS}$ and $F_{\rm C}$ functionals at different days of the simulated growth season. It is clear that $F_{\rm C}$ is both more sensitive to the changes in potatoes and has a ``deeper'' minimum that is easier to detect. 

\begin{figure}
\begin{center}
\includegraphics[width=13cm]{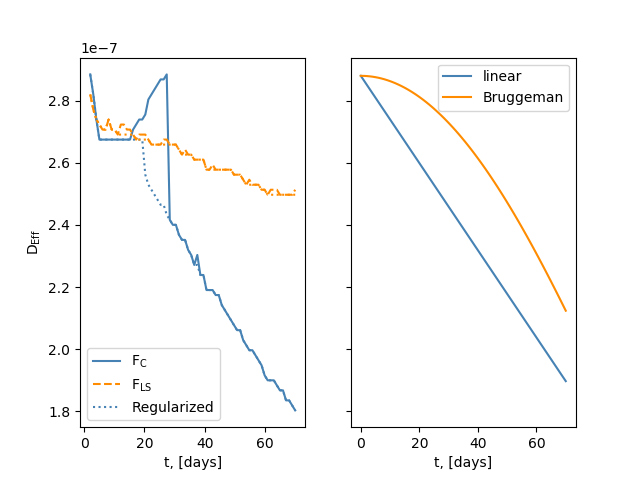}
\end{center}
\caption{\label{Fig7} Left: effective diffusivity $D_{\rm eff}(t)$ as a function of time in days during the growth season reconstructed by minimizing $F_{\rm LS}$ and $F_{\rm C}$ cost functionals over $24$~hour intervals. Dotted lines show the regularized results. Right: analytical approximations based on the mixing formulas (\ref{eq:DeffBrg}) and (\ref{eq:DeffC}).}
\end{figure}

Figure~\ref{Fig7} shows the evolution of $D_{\rm eff}$ throughout the season, where the effective diffusivity has been determined by finding the global minima of $F_{\rm LS}[D]$ and $F_{\rm C}[D]$ (solid lines). To this end the effective model has been run for a range of $100$ values of the diffusivity $D$ uniformly distributed  between $D_{\rm p}$ and $D_{\rm s}$, and the corresponding temperature data were saved and reused to produce functionals $F_{\rm LS/C}$ for each day of the season. There exist much more efficient reduced-order approaches to this kind of problems, e.g. \cite{budko_electromagnetic_2004}, which are not discussed in the present proof of the principle study. 

The functional $F_{C}$ is clearly not unimodal and its second minimum happens within the interval of interest. Moreover, sometimes the minimum that corresponds to physically acceptable effective diffusivity is not a global minimum. Thus, finding a global minimum is not necessarily a good strategy in effective inversion. In Fig.~\ref{Fig7}, $D_{\rm eff}(t)$ at the global minimum is shown with a solid line. While it features an almost linear decline with time after day $30$, prior to that it demonstrates sporadic growth, which is caused by the right minimum in Fig.~\ref{Fig6} being somewhat deeper than the left one. Fortunately, in the present case it is quite easy to decide which of the two minima to choose by incorporating the {\em a priori} dynamic information that potatoes grow rather than become smaller with time. This means that whenever $D_{\rm eff}(t)$ tends to increase one should choose the second minimum of the functional. In cases where the second minimum can not be detected, the value of $D_{\rm eff}(t)$ obtained at the previous day is retained. The result of this simple regularization strategy is shown in Fig.~\ref{Fig7} with dotted lines.

Computing an effective constitutive parameter for a medium consisting of inclusions in a homogeneous host material is the subject of effective medium theory \cite{choy_effective_1999}. According to the steady-state Bruggeman theory the effective parameter depends on the volume fraction of inclusions. In the present two-dimensional case $D_{\rm eff}$ is expected to satisfy the following equation:
\begin{align}
\label{eq:Brg}
\tilde{A_{\rm p}}\,\frac{D_{\rm p}-D_{\rm eff}}{D_{\rm p}+D_{\rm eff}} + 
\left(1-\tilde{A}_{\rm p}\right)\,\frac{D_{\rm s}-D_{\rm eff}}{D_{\rm s}+D_{\rm eff}}=0,
\end{align}
where $\tilde{A}_{\rm p}(t)=A_{\rm p}(t)/A_{\rm eff}$, $\tilde{A}_{\rm p}(0)=0$, $\tilde{A}_{\rm p}(t=70\,\text{days})=0.382$ is the relative area of inclusions (potatoes) at day $t$ and $A_{\rm eff}=0.355\,\text{m}^{2}$ is the area of the effective domain depicted in Fig.~\ref{Fig5} (right). Under the natural assumption that $D_{\rm p}$, $D_{\rm s}$, $D_{\rm eff}$, and $\tilde{A}_{\rm p}$ are all positive the Bruggeman equation simplifies to
\begin{align}
\label{eq:BrgSimpl}
D_{\rm eff}^{2}-(2\tilde{A}_{\rm p}-1)(D_{\rm p}-D_{\rm s})D_{\rm eff}-D_{\rm p}D_{\rm s}=0,
\end{align}
and admits the explicit solution:
\begin{align}
\label{eq:DeffBrg}
D_{\rm eff}(t) = \frac{1}{2}\left[(2\tilde{A}_{\rm p}(t)-1)(D_{\rm p}-D_{\rm s})+
\sqrt{(2\tilde{A}_{\rm p}(t)-1)^{2}(D_{\rm p}-D_{\rm s})^{2}+4D_{\rm s}D_{\rm p}}\right].
\end{align}
Figure~\ref{Fig7}~(right) shows that neither the least-squares nor the correlation-based functional produces $D_{\rm eff}$ that follows Bruggeman's law. The fact that $F_{\rm LS}[D]$ shows a lack of decrease in effective diffusivity with the growth of low-diffusivity inclusions in a high-diffusivity host medium is difficult to explain. Of course, the present configuration, characterized by small particle numbers in a finite and relatively small host domain, makes the applicability of effective medium theories highly questionable. In addition, the $24$-hour time interval used in the daily minimization problems may be too short to capture the instantaneous properties of growing tubers and the heat transfer on a given day is influenced by the state of tubers and temperature distribution of the previous day.

The correlation-based functional $F_{\rm C}[D]$ is, on the other hand, much more sensitive to the tuber size and leads to $D_{\rm eff}^{\rm C}(t)$ which is well approximated by the following simple linear mixing formula:
\begin{align}
\label{eq:DeffC}
D_{\rm eff}^{\rm C}(t) = D_{\rm p}\sqrt{\tilde{A}_{\rm p}(t)} + D_{\rm s}\left(1-\sqrt{\tilde{A}_{\rm p}(t)}\right).
\end{align}
Notice that the effective diffusivity is determined here by the relative ``diameter'' $\sqrt{\tilde{A}_{\rm p}}$ of the potato domain rather than its relative area $\tilde{A}_{\rm p}$. Subsequently, this expression can be used to infer the temporal evolution of $\tilde{A}_{\rm p}(t)$ from $D_{\rm eff}(t)$ as
\begin{align}
\label{eq:Ap}
\tilde{A}_{\rm p}(t) = \left(\frac{D_{\rm eff}^{C}(t)-D_{\rm s}}{D_{\rm p}-D_{\rm s}}\right)^{2}.
\end{align}

\section{Conclusions}
Numerical experiments presented in this paper show that diurnal temperature waves may be used to visualize the growth of tubers, such as potatoes. To this end a sparse array of thermal sensors can be placed below the seed tubers at the time of planting and monitored throughout the season at hourly intervals. While the amplitude of relevant temperature variations is expected to be in the order of $0.1$ degrees, an imaging algorithm is proposed that relies on the relative synchronization rather than absolute precision of sensors. An effective inversion algorithm describing growing tubers in terms of effective time-dependent diffusivity has been demonstrated as well. The effective diffusivity reconstructed from time-domain data depends on the choice of the cost functional. It has been shown that neither the least-squares nor the correlation-based functional leads to an effective diffusivity that follows Bruggeman's mixing formula. The correlation-based functional is more sensitive to the changes in the relative volume of potatoes, but requires additional regularization due to the presence of the second minimum. A simple linear mixing formula is proposed that allows to reconstruct the relative volume of potatoes from the effective diffusivity obtained by minimizing the correlation-based cost functional. 

\section*{References}
\providecommand{\newblock}{}

\end{document}